\documentclass[twocolumn,aps,prl,10pt,showpacs]{revtex4-1}
\usepackage{bm}
\usepackage{amsfonts}
\usepackage{amssymb}
\usepackage{amsmath}
\usepackage{graphicx}
\begin{document}

\title{Time traps for electron-positron pairs}
\author{Iwo Bialynicki-Birula}\email{birula@cft.edu.pl}
\affiliation{Center for Theoretical Physics, Polish Academy of Sciences\\
Aleja Lotnik\'ow 32/46, 02-668 Warsaw, Poland}
\author{Zofia Bialynicka-Birula}
\affiliation{Institute of Physics, Polish Academy of Sciences\\
Aleja Lotnik\'ow 32/46, 02-668 Warsaw, Poland}

\begin{abstract}
Analytical solutions of the Dirac equation in an external electromagnetic field are found such that according to the field-theoretic interpretation  electron-positron pairs are trapped for a period of time. The naive one-particle interpretation of the Dirac wave function fails in this case completely. Simple electromagnetic field which produces this effect was undeniably concocted and may look artificial but the phenomenon of time traps seems real.
\end{abstract}
\maketitle

\section{Introduction}

In the present work we pose and answer the question: Are there configurations of the electromagnetic field that produce the time traps for electron-positron pairs? The time trap is defined here as a time period (see Fig.~1) during which pairs exist, while there is nothing before and after this period. To find the answer we use a simplified theory based on the Dirac equation in an external electromagnetic field, disregarding the mutual interaction of electrons and positrons. This simplification allowed us to state our problem in terms of the properties of the Dirac wave functions.

The Dirac equation, and also other relativistic wave equations, have two levels of interpretation. On one hand they can be treated as wave equations describing the time evolution of the relativistic wave functions. This interpretation, however, must be treated with great caution. A careless approach leads to paradoxes like, for example, the Klein paradox. Solutions of the Dirac equation, in general, are not just wave functions describing the states of electrons but they have also the equally important second part representing (complex conjugate) wave function of the positron. There is no difficulty in separating these two parts in the absence of external field even for very elaborate solutions of the Dirac equation. Also in static electromagnetic fields these two parts can be separated due to the energy gap between the electronic and positronic states. However, when the field is time dependent such separation is not possible since the electron-positron pairs are continuously created and annihilated. The correct interpretation of the wave functions satisfying the Dirac equation was given by Feynman \cite{rpf}. According to this interpretation a solution of the Dirac equation in an external potential describes, in general, four distinct processes depicted in Fig.~2. Still, most authors of papers dealing with the electrons moving in time-dependent electromagnetic fields, starting with the classic paper by Wolkow \cite{wolk}, completely disregard this problem and treat the modulus squared of the Dirac wave function as a true probability density for the electrons only. In the case of time traps, this interpretation fails completely but we can still use the Dirac wave function as a convenient mathematical tool provided we apply their field-theoretic interpretation as we have done in this work.
\begin{figure}
\includegraphics[width=0.4\textwidth,height=0.2\textheight]{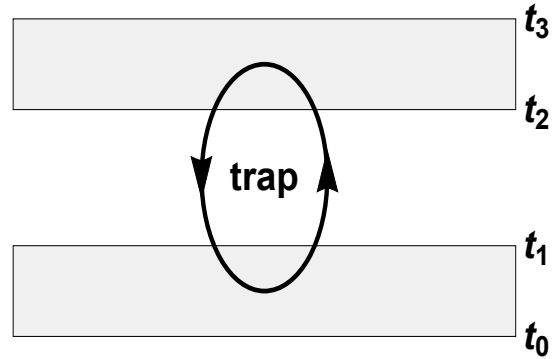}
\caption{The time trap for electron-positron pairs. For the time values before $t_0$ and after $t_3$ the system is in the vacuum state.}\label{Fig1}
\end{figure}
\begin{figure}
\vspace{0.3cm}
\includegraphics[width=0.4\textwidth,height=0.2\textheight]{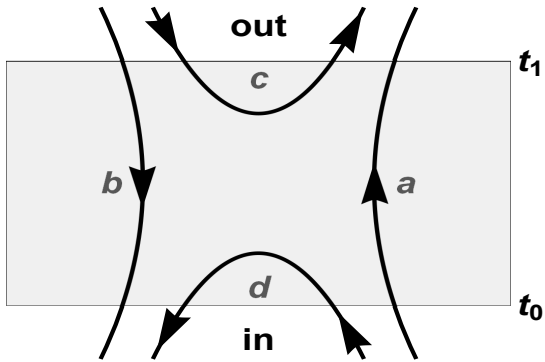}
\caption{Four different processes described by a single solution of the Dirac equation: a) Electron scattering, b) Positron scattering, c) Pair creation, d) Pair annihilation. The shaded area represents the presence of the external electromagnetic field which is responsible for all four processes.}\label{Fig2}
\end{figure}

\section{Solutions of the Dirac equation and the S-matrix}

In the situation depicted in Fig.~1 the electromagnetic field vanishes in the past and in the future so that the formalism of the S-matrix is most convenient to deal with the time evolution. In this formalism the {\em initial/final} state of the system is described by the state vectors in the {\em in/out} Fock spaces build by the action of the {\em in/out} creation operators of noninteracting particles on the {\em in/out} vacuum state vectors. The relevant dynamical properties in this formalism are contained in the S-matrix which transforms the {\em in} operators into the {\em out} operators and the {\em out} vacuum state vector into the {\em in} vacuum state vector,
\begin{align}\label{inout}
a_{out}=S^\dagger a_{in}S,\;\;a^\dagger_{out}=S^\dagger a^\dagger_{in}S,\;\;
S|0_{out}\rangle=|0_{in}\rangle.
\end{align}
In the standard formulation of QED the $S$-matrix is defined as the time-ordered exponential of the interaction Hamiltonian. Given the $S$-matrix, we may find the relation between the {\em out} and {\em in} operators. In this work, however, we reverse this order. First, we find the relation between the {\em out} and {\em in} operators and in the next step we determine the $S$-matrix. The relation between the {\em out} and {\em in} operators can be found directly when an explicit solution of the Dirac equation for the field operator $\hat{\psi}$ is known. This is a hopeless task in the full-fledged theory. However, this task becomes feasible in the simplified version of the theory when only the external classical electromagnetic field is taken into account. In this case the wave function and the field operator both obey the same equation. We may construct the field operator by simply replacing the c-number amplitudes by the annihilation and creation operators.

In the {\em in} and {\em out} regions where the electromagnetic field vanishes the field operator $\hat{\psi}$ can be represented as the standard superposition of annihilation operators of electrons and creation operators of positrons for the free field ($c=1,\hbar=1$),
\widetext
\begin{align}\label{pw}
&\hat{\psi}({\bm r},t)=\sum_{s=\pm}\int\!\frac{d^3p}{(2\pi)^{3/2}}e^{i{\bm p}\cdot{\bm r}}\left[a({\bm p},s)u({\bm p},s)e^{-iE_pt}+b^\dagger(-{\bm p},s)v(-{\bm p},s)e^{iE_pt}\right].
\end{align}
The bispinors $u$ and $v$ will be chosen in the Weyl representation \cite{weyl} of $\gamma$ matrices,
\begin{align}\label{uv}
u({\bm p},+)&=N(\bm p)\left[\!\begin{array}{c}
K(\bm p)+p_z\\p_x+i p_y\\K(\bm p)-p_z\\-p_x-i p_y
\end{array}\right],\;
v({\bm p},+)=N(\bm p)\left[\begin{array}{c}
-K(\bm p)+p_z\\p_x+i p_y\\K(\bm p)+p_z\\p_x+i p_y
\end{array}\right],\quad K(\bm p)=E(\bm p)+m,\nonumber\\
u({\bm p},-)&=N(\bm p)\left[\begin{array}{c}
p_x-i p_y\\K(\bm p)-p_z\\-p_x+ i p_y\\K(\bm p)+p_z
\end{array}\right],\;
v({\bm p},-)=N(\bm p)\left[\begin{array}{c}
p_x-i p_y\\-K(\bm p)-p_z\\p_x-i p_y\\K(\bm p)-p_z
\end{array}\right],\quad N(\bm p)=\frac{1}{2\sqrt{K(\bm p)E(\bm p)}}.
\end{align}
\endwidetext
The time evolution of $\hat{\psi}({\bm r},t)$ from $t_{\rm in}$ to $t_{\rm out}$, when both $t_{\rm in}$ and $t_{\rm out}$ lie in the field free regions (Fig.~2), induces the transformation (\ref{inout}) of the annihilation and creation operators and determines the $S$ operator. In the case of the interaction of electrons with the classical electromagnetic field only, this transformation is a linear one.

\section{Toy model of the electromagnetic field}

Our greatly oversimplified model of the electromagnetic field will be assumed as homogeneous in space, ${\bm E}(t)=-\partial_t{\bm A}(t)$. In addition, the potential ${\bm A}(t)$ will be assumed to be piecewise constant in time. Thus, the electric field consists of $\delta$-like spikes at times $t_n$ when the potential is discontinuous. Owing to the homogeneity of the field, different momentum modes are not coupled and we may consider just one momentum mode at a time. The time dependence of the potential requires the presence of the positive and negative energy components. The field operator of a single momentum mode (but two spin modes) in the $n$-th time slice is:
\begin{align}\label{mode}
&\quad\hat{\psi}_n(\bm r,t)
=e^{i{\bm p}\cdot{\bm r}}\sum_{s=\pm}\\
&\!\times\left[a_n(s)u({\bm p}_n,s)e^{-iE({\bm p}_n)t}+b^\dagger_n(s) v(-{\bm p}_n,s)e^{iE({\bm p}_n)t}\right].\nonumber
\end{align}
It satisfies the solution of the Dirac equation in our piecewise potential,
\begin{align}\label{deq}
(i\gamma^\mu\partial_\mu-\gamma^\mu eA_\mu(n)-m)\hat{\psi}_n(\bm r,t)=0
\end{align}
provided the momentum in the $n$-th time slice is shifted ${\bm p}_n={\bm p}+{\bm q}_n$ by the value of momentum  ${\bm q}_n=e{\bm A}(n)$ delivered by the electric field. However, in order to complete the construction of the Dirac bispinor, we must secure its continuity for all transition times $t_n$. This can be achieved by properly adjusting the annihilation and creation operators in the adjacent time slices.

Owing to the orthogonality of the bispinors $u$ and $v$, we may extract the operators $(a_{n+1},b^\dagger_{n+1})$ just by multiplying the continuity condition
\begin{align}\label{cont}
\hat{\psi}_{n+1}(\bm r,t_n)=\hat{\psi}_n(\bm r,t_n),
\end{align}
by $u^\dagger({\bm p}_{n+1},s)$ and $v^\dagger({\bm p}_{n+1},s)$. The resulting relations can be written in the following compact matrix form,
\begin{align}\label{mat}
P_{n+1}(t_n)F_{n+1}=M_nP_n(t_n)F_{n},
\end{align}
where the transfer matrices $M_n$ are built from all 16 products of the bispinors $u$ and $v$,
\begin{align}\label{etc}
M_n^{11}&=u^\dagger({\bm p}_{n+1},+)u({\bm p}_n,+),\nonumber\\
M_n^{12}&=u^\dagger({\bm p}_{n+1},+)u({\bm p}_n,-),\nonumber\\
M_n^{13}&=u^\dagger({\bm p}_{n+1},+)v({\bm p}_n,+),\;{\rm etc.},
\end{align}
the four operators were arranged in a vector, and $P_n(t_n)$ is the diagonal time evolution matrix,
\begin{align}\label{fp}
F_n&=\left[a_n(+),a_n(-),b^\dagger_n(+),b^\dagger_n(-)\right],\\
P_n(t)&={\rm diag}\left[e^{-iE({\bf p}_n)t},e^{-iE({\bf p}_n)t},e^{iE({\bf p}_n)t},e^{iE({\bf p}_n)t}\right].\nonumber
\end{align}
Using repeatedly the relation (\ref{mat}) we obtain the formula which expresses the final operators in terms of the initial operators,
\widetext
\begin{align}\label{matn}
P_4(t_3)F_4=M_3P_3(t_3-t_2)M_2P_2(t_2-t_1)M_1P_1(t_1-t_0)M_0P_0(t_0)F_0.
\end{align}
\endwidetext
This is a characteristic formula for any layered medium (cf., for example, Eq.(20) in \cite{soto}) in which the product of transfer matrices connects initial and final layers.
\begin{figure}
\vspace{0.3cm}
\includegraphics[width=0.4\textwidth,height=0.2\textheight]
{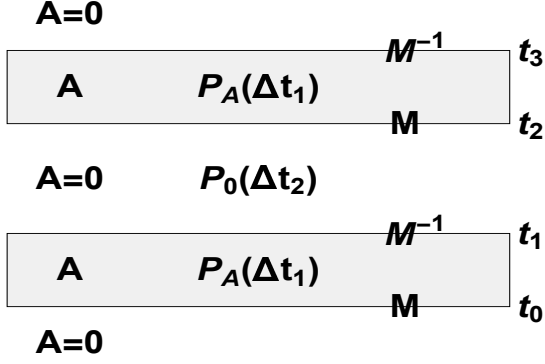}
\caption{Schematic representation of the time trap ($\Delta t_1=t_1-t_0=t_3-t_2, \Delta t_2=t_2-t_1$). Note the symmetry under time reversal. Only one transfer matrix $M$ and two propagation matrices $P$ are needed to obtain the complete description.}\label{Fig3}
\end{figure}

In what follows we will consider only the simplest case, when there is no coupling between different spin state. This happens when the momentum is in the direction of the electric field, $p_x=0=p_y$. The transformations between the operators $a({\bm p},s)$ and $b^\dagger(-{\bm p},s)$ (known in theory of superconductivity as Bogoljubov-Valatin transformations \cite{bog,val}) are characterized by $2\times 2$ matrices. For definiteness, we consider the spin $+$ components. The continuity condition at time $t_0$ in Fig.~3 reads,
\begin{align}\label{mm1}
\left[\begin{array}{c}a_1\\
b^\dagger_1\end{array}\right]
=\left[\begin{array}{cc}m_{11}&m_{12}\\
m_{21}&m_{22}\end{array}\right]
\left[\begin{array}{c}a_0\\
b^\dagger_0\end{array}\right],
\end{align}
where
\begin{subequations}
\begin{align}\label{mm2}
m_{11}=m_{22}=\frac{K(p+q)K(p)+p(p+q)}
{2\sqrt{E(p+q)E(p)K(p+q)K(p)}},\\
m_{12}=-m_{21}=\frac{pE(p+q)-(p+q)E(p)-mq}
{2\sqrt{E(p+q)E(p)K(p+q)K(p)}}.
\end{align}
\end{subequations}
We dropped the vector indices because both ${\bm p}$ and ${\bm q}$ have only the $z$ component. The matrix $m_{ij}$ is unitary, as it must be, to preserve the anticommutation relations between the annihilation and creation operators. The product of the first four factors in (\ref{matn}) gives the transformation of the operators at time $t_0$ into the operators at time $t_1$. The properties of this matrix are crucial for the formation of the time trap.

The transformations (\ref{mm1}) are generated by c-number unitary matrices. However, these transformations can also be implemented by unitary operators acting in the Fock space. In our simple case the equivalence of the two forms of the transformation gives:
\begin{align}\label{bv}
\left[\begin{array}{cc}u_{11}&u_{12}\\
u_{21}&u_{22}\end{array}\right]
\left[\begin{array}{c}a\\
b^\dagger\end{array}\right]={\hat U}^\dagger\left[\begin{array}{c}a\\
b^\dagger\end{array}\right]{\hat U}.
\end{align}
The mathematical description in terms of simple $2\times 2$ matrices is much easier to handle while the description of the same situation in terms of unitary operators ${\hat U}$ provides the physical interpretation of the results.

A general unitary $2\times 2$ matrix can be parametrized with 4 real coefficients,
\begin{align}\label{u2}
\left[\begin{array}{cc}u_{11}&u_{12}\\
u_{21}&u_{22}\end{array}\right]=e^{i\xi_0}
\left[\cos(|\bm\xi|)-i{\bm\xi}\!\cdot\!{\bm\sigma}
\frac{\sin(|\bm\xi|)}{|\bm\xi|}\right],
\end{align}
where ${\bm\xi}=\{\xi_1,\xi_2,\xi_3\}$ and $\{\sigma_1,\sigma_2,\sigma_3\}$ are the Pauli matrices. The corresponding unitary operator ${\hat U}$ in (\ref{bv}) is:
\begin{align}\label{u3}
{\hat U}=\exp[i(\xi_0G_0+{\bm\xi}\cdot{\bm G})].
\end{align}
The four generators $G_0$ and ${\bm G}$ are:
\begin{align}\label{u4}
&G_0=a^\dagger a-b^\dagger b,\quad\;\;\;\,
G_1=a b+b^\dagger a^\dagger,\nonumber\\
&G_2=i(a b-b^\dagger a^\dagger),\quad G_3=-a^\dagger a-b^\dagger b.
\end{align}
The fermionic nature of the annihilation and creation operators enables one to express the exponential operator (\ref{u3}) as the following combination of the annihilation and creation operators,
\begin{align}\label{u5}
&e^{i\xi_3}{\hat U}=\left[\cos(|\bm\xi|)+i\xi_3
\frac{\sin(|\bm\xi|)}{|\bm\xi|}\right](I-a^\dagger a-b^\dagger b)\nonumber\\
&+[e^{i\xi_0}a^\dagger a+e^{-i\xi_0}b^\dagger b]+2[\cos(|\bm\xi|)-\cos(\xi_0)]a^\dagger b^\dagger b a\nonumber\\
&+i\frac{\sin(|\bm\xi|)}{|\bm\xi|}\left[(\xi_1+i\xi_2)a b-(\xi_1-i\xi_2)a^\dagger b^\dagger\right],
\end{align}
where $I$ is the unit operator. In order to construct the time trap, we need to apply the transition formulas (\ref{bv}) for consecutive time slices at each time $t_n$. The resulting chain of transformations can be evaluated either in terms of the products of bispinors $u$ and $v$ or equivalently in terms of the products of the operators $\hat U$. We choose the first method because the formula (\ref{u5}) is rather complicated and then transcribe the final results into the formalism of quantum states.

\section{Construction of the time trap}

We define the time trap for the field configuration depicted in Fig.~1 when the state of the system is such that it is the vacuum state for the times before $t_0$ and after $t_3$ but it is the state of electron-positron pairs during the time period from $t_1$ to $t_2$. Such a state can be achieved by fine-tuning the values of two parameters: the momentum and time duration. The criterion for the time trap will be formulated in terms of the properties of the $u_{ij}$ matrix.

Let us consider the transfer matrix that connects the annihilation and creation operators at times $t_0$ and $t_1$. In order to form the time trap we require that the vacuum state at $t_0$ will be transformed into the pair state at $t_1$. The matrix $u_{ij}$ which realizes this transformation must satisfy the conditions $u_{11}=0=u_{22}$. The proof of this assertion is based on (\ref{bv}) rewritten in the form:
\begin{subequations}
\begin{align}\label{sm}
a^\dagger {\hat U}|0\rangle={\hat U}(u^*_{11}a^\dagger+u^*_{12}b)|0\rangle,\\
b^\dagger{\hat U}|0\rangle={\hat U}(u_{21}a+u_{22}b^\dagger)|0\rangle.
\end{align}
\end{subequations}
Under the assumption that ${\hat U}|0\rangle$ is the pair state, the left hand sides in these formulas must vanish because one cannot add more particles to the state containing the pair. The vanishing of the right hand sides requires that $u_{11}=0=u_{22}$. The product of matrices $u_B=MP(t_1-t_0)M^{-1}$  that evolves our system across the ``potential barrier'' (see Fig.~3) has the general form (\ref{bv}) with the following values of the parameters:
\begin{align}\label{up}
&|{\bm\xi}|=E(p+q)(t_1-t_0),\\
\frac{\xi_1}{|{\bm\xi}|}=-&\frac{m q}{E(p)E(p+q)}\quad
\frac{\xi_3}{|{\bm\xi}|}=\frac{m^2+p^2+p q}{E(p)E(p+q)}.
\end{align}
The only way to make the diagonal elements of this matrix vanish is to choose the time difference such that $|{\bm\xi}|=(n+1/2)\pi$ and the momentum $p=(-q\pm\sqrt{q^2-4m^2})$. Note that the value of $p$ is real only when the electric field is strong enough to produce pairs, $eA\ge 2m$. With this choice of parameters the matrix $u_B$ takes on the form:
\begin{align}\label{up1}
u_B=\left[\begin{array}{cc}0&i\\i&0
\end{array}\right],
\end{align}
The assumed time symmetry gives the same matrix for the propagation from time $t_2$ to time $t_3$. To complete the construction of the matrix $u_T$ that generates the time trap we must insert the free propagation matrix from time $t_1$ to time $t_2$  between the two matrices $u_B$,
\begin{align}\label{ut}
u_T=\left[\begin{array}{cc}0&i\\i&0\end{array}\right]
\left[\begin{array}{cc}e^{-iE({\bf p}_n)(t_2-t_1)}&0\\0&\hspace{-0.8cm}e^{iE({\bf p}_n)(t_2-t_1)}\end{array}\right]
\left[\begin{array}{cc}0&i\\i&0\end{array}\right].
\end{align}
Choosing the time period $E(p)(t_2-t_1)=(2n+1)\pi$, we obtain the unit matrix $u_T=I$. The description of the time trap is very simple in terms of the unitary operators ${\hat U}_T$ and ${\hat U}_B$ in the Fock space corresponding to the matrices $u_T$ and $u_B$. The operator ${\hat U}_T$ does noting because it is the unit operator. The trap is completely transparent,it has no influence on particles evolving in time from $t_0$ to $t_3$. The operator ${\hat U}_B$ does not change the one-particles states but it converts the vacuum state into the pair state, and vice versa. If the initial state is the vacuum, then we encounter pairs in time trap but there is again the vacuum in the final state.

The symmetry between the vacuum state and the pair state makes it possible to produce time crystals by repeating periodically the trap configuration. Of course, this periodic trap is manufactured rather than created spontaneously, so it is not the kind of the time crystal envisioned by Wilczek \cite{fw}.

\section{Discussion}

We have to admit that the model of the time trap described here is painfully unrealistic. Its only value is its simplicity. All calculations can be explicitly carried out so that we could prove the main point of our analysis that the study of the solutions of the Dirac equation in an external electromagnetic field may lead to results extending much beyond the mere description of bound states of electrons in atoms and relativistic beams of electrons. Note that our solutions of the Dirac equation are still valid if the annihilation operators were replaced by complex amplitudes, say $f_n$ and $g^*_n$ obeying the relations imposed by the continuity conditions. In this case instead of the field operator we would obtain a solution of the Dirac equation. Our results clearly show that one c-number solution of the Dirac equation may describe a fairly complicated history. Even though the calculations may be carried out with the use of the Dirac wave function, it is difficult to obtain the correct physical interpretation of such solutions without resorting to the field-theoretic tools.

\end{document}